\documentclass[onecolumn]{article}
\usepackage{graphicx}
\usepackage[utf8]{inputenc}
\usepackage{amsmath}
\usepackage{physics}
\usepackage{mathtools}
\usepackage{amssymb}
\usepackage{slashed}
\usepackage{tikz-feynman}
\usepackage{amsthm}
\usepackage{txfonts}
\usepackage{geometry}
\begin{document}

\title{Neutrino Masses and Noncyclic Geometric Phase of Entangled Charged Lepton-Neutrino Pair in External Magnetic Field}
\author{Lu Jianlong\\ \emph{\small{Department of Physics, National University of Singapore}}}
\date{}
\maketitle

\abstract
In this paper, we consider a pair of entangled charged lepton-neutrino embedded in a constant magnetic field, which can be produced in the purely leptonic decay of charged pseudoscalar mesons. Particles with nonzero magnetic dipole moment will precess in the present of external magnetic field and induce a nontrivial noncyclic geometric phase. The dependence of this geometric phase on neutrino mass can be used to determine the Dirac/Majorana nature of massive neutrinos.

\section{INTRODUCTION}
The concept of neutrino originated from Wolfgang Pauli's proposal in 1930 \cite{pauli}, which was aiming to solve the problem of continuous energy spectrum of electrons produced in beta decay. In the present Standard Model of particle physics, neutrinos (and antineutrinos) are modeled as massless elementary fermions. However, the observed flavor-changing phenomena of neutrinos strongly suggest, although not necessarily \cite{lu}, that neutrinos should have nonzero nondegenerate masses such that their flavor eigenstates and mass eigenstates mix. The mismatch between these two groups of eigenstates is packaged in the Pontecorvo–Maki–Nakagawa–Sakata matrix (PMNS matrix) $U$ as follows
\begin{equation}
   \begin{pmatrix}\ket{\nu_{e}} \\ \ket{\nu_{\mu}}\\ \ket{\nu_{\tau}}\end{pmatrix}=U\begin{pmatrix}\ket{\nu_{1}}\\ \ket{\nu_{2}}\\ \ket{\nu_{3}}\end{pmatrix}
\end{equation}
where we have assumed that the number of neutrino flavors is three, as supported by the decay width of Z bosons \cite{zboson}\cite{PDG}. The standard parametrization of PMNS matrix adopted by Particle Data Group is \cite{PDG}\cite{XZZ}
\begin{equation}
    U= \begin{pmatrix} c_{12}c_{13} & s_{12}c_{13} & s_{13}e^{-i\delta}\\ -s_{12}c_{23}- c_{12}s_{23}s_{13} e^{i\delta} & c_{12}c_{23}-s_{12}s_{23}s_{13}e^{i\delta} & s_{23}c_{13} \\ s_{12}s_{23}-c_{12}c_{23}s_{13}e^{i\delta} & -c_{12}s_{23}-s_{12}c_{23}s_{13}e^{i\delta} & c_{23}c_{13}\end{pmatrix}\begin{pmatrix}e^{i\phi_{1}} & & \\  & e^{i\phi_{2}} & \\ & & 1 \end{pmatrix}
\end{equation}
in which totally six parameters appear, including three neutrino mixing angles $(\theta_{12},\theta_{13},\theta_{23})$ and three CP-violation phases $(\delta,\phi_{1},\phi_{2})$. $\delta$ is called Dirac CP-violating phase, which is responsible for those CP-violating effects in neutrino oscillations. $\phi_{1}$ and $\phi_{2}$ are called Majorana CP-violating phases. Majorana CP-violating phases are irrelevant in neutrino oscillations \cite{giu}, and are physically significant only if the massive neutrinos are Majorana particles. In the Lagrangian density of the Standard Model, no neutrino mass terms exist, corresponding to the massless neutrino assumption. By appropriately modifying the original Standard Model, there are three types of neutrino mass terms we can introduce into the Lagrangian density, including Dirac mass term, Majorana mass term and Dirac+Majorana mass term. Dirac mass term corresponds to the Dirac nature of massive neutrinos, while both Majorana mass term and Dirac+Majorana mass term imply that massive neutrinos are Majorana particles. Although the neutrino magnetic moment vanishes in the cases of massless neutrinos and massive Majorana neutrinos, a massive Dirac neutrino with mass $m_{\nu}$ can have a nonzero magnetic dipole moment \cite{fujikawa}
\begin{equation}
   \mu_{\nu}=\frac{3eG_{F}m_{\nu}}{8\sqrt{2}\pi^{2}}
\end{equation}
where $e$ is the absolute value of the fundamental electric charge, $G_{F}$ is Fermi's coupling constant.\\
A particle with nonzero magnetic dipole moment will precess in an external magnetic field, with angular frequency proportional to the strength of magnetic field. The precession induces a nontrivial noncyclic geometric phase both in the case of single particle and in the case of entangled particle pair. \cite{sjo} This noncyclic geometric phase is defined as the difference between total phase and dynamical phase, i.e.,
\begin{equation}
    \Phi_{G}=\Phi_{T}-\Phi_{D}={\rm arg}\bra{\Psi(0)}\ket{\Psi(\tau)}+i\int_{0}^{\tau}\bra{\Psi(t)}\ket{\dot{\Psi}(t)}dt.
\end{equation}
In the following section, we calculate the noncyclic geometric phase of an entangled pair of particles consisting of one charged lepton and one neutrino in an external magnetic field. The reduced Planck constant $\hbar$ is normalized to $1$. The Dirac/Majorana nature-dependence of geometric phase of neutrinos have been discussed, for example, in \cite{capolupo}. Our paper proceeds in a different direction, generalizing the results in \cite{sjo} to an entangled charged lepton-neutrino pair in which the Larmor frequency of the neutrino subsystem is Dirac/Majorana nature-dependent and does not have a definite value.

\section{CALCULATION}
Schmidt's theorem ensures that we can always decompose the state of a pair of spin $\frac{1}{2}$ particles as \cite{schmidt} \cite{peres}
\begin{equation}
   \ket{\Psi}=e^{-i\frac{\beta}{2}}\cos\frac{\alpha}{2}\ket{\textbf{n}}\ket{\textbf{m}}+ e^{i\frac{\beta}{2}}\sin\frac{\alpha}{2}\ket{-\textbf{n}}\ket{-\textbf{m}},
\end{equation}
in which we have two pair of antipodal points on the Poincaré sphere:
\begin{equation}
\begin{cases}
   \ket{\textbf{n}}= e^{-i\frac{\phi_{1}}{2}}\cos\frac{\theta_{1}}{2}\ket{+z}+ e^{i\frac{\phi_{1}}{2}}\sin\frac{\theta_{1}}{2}\ket{-z}\\
   \ket{-\textbf{n}}= -i e^{-i\frac{\phi_{1}}{2}}\sin\frac{\theta_{1}}{2}\ket{+z}+ i e^{i\frac{\phi_{1}}{2}}\cos\frac{\theta_{1}}{2}\ket{-z}
\end{cases},
\end{equation}
\begin{equation}
\begin{cases}
   \ket{\textbf{m}}= e^{-i\frac{\phi_{2}}{2}}\cos\frac{\theta_{2}}{2}\ket{+z}+ e^{i\frac{\phi_{2}}{2}}\sin\frac{\theta_{2}}{2}\ket{-z}\\
   \ket{-\textbf{m}}= -i e^{-i\frac{\phi_{2}}{2}}\sin\frac{\theta_{2}}{2}\ket{+z}+ i e^{i\frac{\phi_{2}}{2}}\cos\frac{\theta_{2}}{2}\ket{-z}
\end{cases}.
\end{equation}
Note that the antipodal point pairs satisfy $\bra{-\textbf{n}}\ket{\textbf{n}}=0$ and $\bra{-\textbf{m}}\ket{\textbf{m}}=0$. There are two parameters in the state of the whole system, $\alpha$ and $\beta$. The degree of entanglement between these two particles (subsystems) is indicated by the former. For example, when $\alpha=0$ or $\alpha=\pi$ the state of the whole system reduces to a product state, while $\alpha=\frac{\pi}{2}$ shows maximal entanglement. For entangled particle pair embedded in external constant magnetic field, the time-dependence of $\ket{\pm \textbf{n}}$ and $\ket{\pm \textbf{m}}$ are hidden in the parameters $\phi_{1}$ and $\phi_{2}$,
\begin{equation}
   \phi_{i}=\bar{\phi}_{i}+\omega_{i}t,\ \ \ i=1,2
\end{equation}
where $\omega_{i}$ are Larmor frequencies of corresponding subsystems and $\bar{\phi}_{i}$ are initial values of $\phi_{i}$. The explicit time-dependent form of $\ket{\Psi}$ is
\begin{equation}
\begin{gathered}
   \ket{\Psi(t)}
   = e^{-i\frac{(\bar{\phi}_{1}+\bar{\phi}_{2})+(\omega_{1}+\omega_{2})t}{2}}  \Big(e^{-i\frac{\beta}{2}}\cos\frac{\alpha}{2}\cos\frac{\theta_{1}}{2}\cos\frac{\theta_{2}}{2}- e^{i\frac{\beta}{2}}\sin\frac{\alpha}{2}\sin\frac{\theta_{1}}{2}\sin\frac{\theta_{2}}{2}  \Big) \ket{+z}\ket{+z}\\
      + e^{-i\frac{(\bar{\phi}_{1}-\bar{\phi}_{2})+(\omega_{1}-\omega_{2})t}{2}}    \Big(e^{-i\frac{\beta}{2}}\cos\frac{\alpha}{2}\cos\frac{\theta_{1}}{2}\sin\frac{\theta_{2}}{2}+ e^{i\frac{\beta}{2}}\sin\frac{\alpha}{2}\sin\frac{\theta_{1}}{2}\cos\frac{\theta_{2}}{2}  \Big) \ket{+z}\ket{-z}\\
      + e^{i\frac{(\bar{\phi}_{1}-\bar{\phi}_{2})+(\omega_{1}-\omega_{2})t}{2}}  \Big(e^{-i\frac{\beta}{2}}\cos\frac{\alpha}{2}\sin\frac{\theta_{1}}{2}\cos\frac{\theta_{2}}{2}+ e^{i\frac{\beta}{2}}\sin\frac{\alpha}{2}\cos\frac{\theta_{1}}{2}\sin\frac{\theta_{2}}{2} \Big) \ket{-z}\ket{+z}\\
      + e^{i\frac{(\bar{\phi}_{1}+\bar{\phi}_{2})+(\omega_{1}+\omega_{2})t}{2}}  \Big(e^{-i\frac{\beta}{2}}\cos\frac{\alpha}{2}\sin\frac{\theta_{1}}{2}\sin\frac{\theta_{2}}{2}-e^{i\frac{\beta}{2}}\sin\frac{\alpha}{2}\cos\frac{\theta_{1}}{2}\cos\frac{\theta_{2}}{2} \Big) \ket{-z}\ket{-z}.
\end{gathered}
\end{equation}
During time evolution, the total phase obtained by the state of the whole system is 
\begin{equation}
\begin{gathered}
   {\rm arg} \bra{\Psi(0)}\ket{\Psi(\tau)}
    = -\arctan \Big(\frac{\tan\frac{\omega_{1}\tau}{2}\cos\alpha\cos\theta_{1}+\tan\frac{\omega_{2}\tau}{2}\cos\alpha\cos\theta_{2}  }{1 -\tan\frac{\omega_{1}\tau}{2}\tan\frac{\omega_{2}\tau}{2}\cos\theta_{1}\cos\theta_{2} 
     +\tan\frac{\omega_{1}\tau}{2}\tan\frac{\omega_{2}\tau}{2} \cos\beta \sin\alpha\sin\theta_{1}\sin\theta_{2}  } \Big).
\end{gathered}
\end{equation}
And the accompanying dynamical phase is 
\begin{equation}
\begin{gathered}
    -i\int_{0}^{\tau} \bra{\Psi(t)}\ket{\dot{\Psi}(t)}dt=- \frac{\omega_{1}\tau}{2}\cos\alpha\cos\theta_{1}-\frac{\omega_{2}\tau}{2}\cos\alpha\cos\theta_{2}.
\end{gathered}
\end{equation}
Therefore, the noncyclic geometric phase obtained by the state of the entangled pair during the time evolution from $t=0$ to $t=\tau$ is as follows:
\begin{equation}
\begin{gathered}
    \Phi_{G}[\Psi(t);0,\tau]=  -\arctan \Big(\frac{\tan\frac{\omega_{1}\tau}{2}\cos\alpha\cos\theta_{1}+\tan\frac{\omega_{2}\tau}{2}\cos\alpha\cos\theta_{2}  }{1 -\tan\frac{\omega_{1}\tau}{2}\tan\frac{\omega_{2}\tau}{2}\cos\theta_{1}\cos\theta_{2} 
     +\tan\frac{\omega_{1}\tau}{2}\tan\frac{\omega_{2}\tau}{2} \cos\beta \sin\alpha\sin\theta_{1}\sin\theta_{2}  } \Big)\\ + \frac{\omega_{1}\tau}{2}\cos\alpha\cos\theta_{1}+\frac{\omega_{2}\tau}{2}\cos\alpha\cos\theta_{2}.
\end{gathered}
\end{equation}
If the neutrino in the entangled pair has zero magnetic moment (possibly due to zero mass or nonzero Majorana mass), then it will not precess in the magnetic field, i.e., $\omega_{2}=0$. By directly substituting $\omega_{2}$ into the previous formula of noncyclic geometric phase, we have 
\begin{equation}
\begin{gathered}
    \Phi_{G}[\Psi(t);0,\tau]_{\omega_{2}=0}=  -\arctan \Big(\tan\frac{\omega_{1}\tau}{2}\cos\alpha\cos\theta_{1} \Big)+ \frac{\omega_{1}\tau}{2}\cos\alpha\cos\theta_{1}.
\end{gathered}
\end{equation}
If the neutrino in the entangled pair is in its mass eigenstate, with nonzero Dirac mass $m_{\nu}$, then the corresponding magnetic moment is $\mu_{\nu}=\frac{3eG_{F}m_{\nu}}{8\sqrt{2}\pi^{2}}$ as mentioned previously. This nonzero magnetic moment leads to neutrino spin precession in magnetic field with angular frequency $\omega_{2}$ determined by the neutrino gyromagnetic ratio $\gamma_{\nu}$ and the magnetic field strength $B$, 
\begin{equation}
    \omega_{2}=\gamma_{\nu} B=\frac{\mu_{\nu}}{S_{\nu}}B= \frac{3eG_{F}m_{\nu}B}{4\sqrt{6}\pi^{2}}.
\end{equation}
On the other hand, the angular frequency of the spin precession of the charged lepton is 
\begin{equation}
    \omega_{1}=\gamma_{l} B,
\end{equation}
proportional to the corresponding lepton gyromagnetic ratio $\gamma_{l}$ .\\
Note that the above noncyclic geometric phase formula cannot be directly applied on the entangled charged lepton-neutrino pair produced simultaneously in purely leptonic decay of charged pseudoscalar mesons, such as muon-neutrino pair produced in charged pion decay \cite{PDG}, because in these cases the ejected neutrino is initially in its flavor eigenstate instead of mass eigenstate. The state vectors $\ket{\textbf{m}}$ and $\ket{-\textbf{m}}$ are now superpositions of different mass eigenstates (with corresponding spin directions), i.e.,
\begin{equation}
    \ket{\textbf{m}}=U_{l 1}\ket{\nu_{1},\textbf{m}}+ U_{l 2}\ket{\nu_{2},\textbf{m}}+ U_{l 3}\ket{\nu_{3},\textbf{m}},
\end{equation}
\begin{equation}
    \ket{-\textbf{m}}=U_{l 1}\ket{\nu_{1},-\textbf{m}}+ U_{l 2}\ket{\nu_{2},-\textbf{m}}+ U_{l 3}\ket{\nu_{3},-\textbf{m}},
\end{equation}
where for $i=1,2,3$ we have 
\begin{equation}
    \ket{\nu_{i},\textbf{m}}= e^{-i\frac{\bar{\phi}_{2}+\omega_{2,i}t}{2}}\cos\frac{\theta_{2}}{2}\ket{\nu_{i},+z}+ e^{i\frac{\bar{\phi}_{2}+\omega_{2,i}t}{2}}\sin\frac{\theta_{2}}{2}\ket{\nu_{i},-z},
\end{equation}
\begin{equation}
    \ket{\nu_{i},-\textbf{m}}= -i e^{-i\frac{\bar{\phi}_{2}+\omega_{2,i}t}{2}}\sin\frac{\theta_{2}}{2}\ket{\nu_{i},+z}+ i e^{i\frac{\bar{\phi}_{2}+\omega_{2,i}t}{2}}\cos\frac{\theta_{2}}{2}\ket{\nu_{i},-z},
\end{equation}
with Larmor frequencies for $i$-th Dirac mass eigenstates
\begin{equation}
    \omega_{2,i}= \frac{3eG_{F}m_{\nu_{i}}B}{4\sqrt{6}\pi^{2}}.
\end{equation}
Now the state of the entangled pair is 
\begin{equation}
\begin{gathered}
    \ket{\Psi(t)}
   = e^{-i\frac{(\bar{\phi}_{1}+\bar{\phi}_{2})+(\omega_{1}+\omega_{2,1})t}{2}} U_{l 1}    \Big(e^{-i\frac{\beta}{2}}\cos\frac{\alpha}{2}\cos\frac{\theta_{1}}{2}\cos\frac{\theta_{2}}{2} -  e^{i\frac{\beta}{2}}\sin\frac{\alpha}{2}\sin\frac{\theta_{1}}{2}\sin\frac{\theta_{2}}{2}        \Big)\ket{+z}\ket{\nu_{1},+z} \\ + e^{-i\frac{(\bar{\phi}_{1}-\bar{\phi}_{2})+(\omega_{1}-\omega_{2,1})t}{2}} U_{l 1}      \Big(e^{-i\frac{\beta}{2}}\cos\frac{\alpha}{2}\cos\frac{\theta_{1}}{2}\sin\frac{\theta_{2}}{2} + e^{i\frac{\beta}{2}}\sin\frac{\alpha}{2}\sin\frac{\theta_{1}}{2}\cos\frac{\theta_{2}}{2}        \Big) \ket{+z}\ket{\nu_{1},-z} \\
   + e^{i\frac{(\bar{\phi}_{1}-\bar{\phi}_{2})+(\omega_{1}-\omega_{2,1})t}{2}} U_{l 1}       \Big(e^{-i\frac{\beta}{2}}\cos\frac{\alpha}{2}\sin\frac{\theta_{1}}{2}\cos\frac{\theta_{2}}{2}+ e^{i\frac{\beta}{2}}\sin\frac{\alpha}{2}\cos\frac{\theta_{1}}{2}\sin\frac{\theta_{2}}{2}   \Big)\ket{-z}\ket{\nu_{1},+z} \\    + e^{i\frac{(\bar{\phi}_{1}+\bar{\phi}_{2})+(\omega_{1}+\omega_{2,1})t}{2}} U_{l 1}     \Big(  e^{-i\frac{\beta}{2}}\cos\frac{\alpha}{2}\sin\frac{\theta_{1}}{2}\sin\frac{\theta_{2}}{2} - e^{i\frac{\beta}{2}}\sin\frac{\alpha}{2}\cos\frac{\theta_{1}}{2}\cos\frac{\theta_{2}}{2}      \Big)\ket{-z}\ket{\nu_{1},-z} \\
   + e^{-i\frac{(\bar{\phi}_{1}+\bar{\phi}_{2})+(\omega_{1}+\omega_{2,2})t}{2}} U_{l 2}    \Big(e^{-i\frac{\beta}{2}}\cos\frac{\alpha}{2}\cos\frac{\theta_{1}}{2}\cos\frac{\theta_{2}}{2} -  e^{i\frac{\beta}{2}}\sin\frac{\alpha}{2}\sin\frac{\theta_{1}}{2}\sin\frac{\theta_{2}}{2}        \Big)\ket{+z}\ket{\nu_{2},+z} \\     + e^{-i\frac{(\bar{\phi}_{1}-\bar{\phi}_{2})+(\omega_{1}-\omega_{2,2})t}{2}} U_{l 2}      \Big(e^{-i\frac{\beta}{2}}\cos\frac{\alpha}{2}\cos\frac{\theta_{1}}{2}\sin\frac{\theta_{2}}{2} + e^{i\frac{\beta}{2}}\sin\frac{\alpha}{2}\sin\frac{\theta_{1}}{2}\cos\frac{\theta_{2}}{2}        \Big)\ket{+z}\ket{\nu_{2},-z} \\
   + e^{i\frac{(\bar{\phi}_{1}-\bar{\phi}_{2})+(\omega_{1}-\omega_{2,2})t}{2}} U_{l 2}       \Big(e^{-i\frac{\beta}{2}}\cos\frac{\alpha}{2}\sin\frac{\theta_{1}}{2}\cos\frac{\theta_{2}}{2}+ e^{i\frac{\beta}{2}}\sin\frac{\alpha}{2}\cos\frac{\theta_{1}}{2}\sin\frac{\theta_{2}}{2}   \Big)\ket{-z}\ket{\nu_{2},+z} \\ + e^{i\frac{(\bar{\phi}_{1}+\bar{\phi}_{2})+(\omega_{1}+\omega_{2,2})t}{2}} U_{l 2}     \Big(  e^{-i\frac{\beta}{2}}\cos\frac{\alpha}{2}\sin\frac{\theta_{1}}{2}\sin\frac{\theta_{2}}{2} - e^{i\frac{\beta}{2}}\sin\frac{\alpha}{2}\cos\frac{\theta_{1}}{2}\cos\frac{\theta_{2}}{2}      \Big)\ket{-z}\ket{\nu_{2},-z} \\
   + e^{-i\frac{(\bar{\phi}_{1}+\bar{\phi}_{2})+(\omega_{1}+\omega_{2,3})t}{2}} U_{l 3}    \Big(e^{-i\frac{\beta}{2}}\cos\frac{\alpha}{2}\cos\frac{\theta_{1}}{2}\cos\frac{\theta_{2}}{2} -  e^{i\frac{\beta}{2}}\sin\frac{\alpha}{2}\sin\frac{\theta_{1}}{2}\sin\frac{\theta_{2}}{2}        \Big)\ket{+z}\ket{\nu_{3},+z}\\ + e^{-i\frac{(\bar{\phi}_{1}-\bar{\phi}_{2})+(\omega_{1}-\omega_{2,3})t}{2}} U_{l 3}      \Big(e^{-i\frac{\beta}{2}}\cos\frac{\alpha}{2}\cos\frac{\theta_{1}}{2}\sin\frac{\theta_{2}}{2} + e^{i\frac{\beta}{2}}\sin\frac{\alpha}{2}\sin\frac{\theta_{1}}{2}\cos\frac{\theta_{2}}{2}        \Big)\ket{+z}\ket{\nu_{3},-z} \\
   + e^{i\frac{(\bar{\phi}_{1}-\bar{\phi}_{2})+(\omega_{1}-\omega_{2,3})t}{2}} U_{l 3}       \Big(e^{-i\frac{\beta}{2}}\cos\frac{\alpha}{2}\sin\frac{\theta_{1}}{2}\cos\frac{\theta_{2}}{2}+ e^{i\frac{\beta}{2}}\sin\frac{\alpha}{2}\cos\frac{\theta_{1}}{2}\sin\frac{\theta_{2}}{2}   \Big)\ket{-z}\ket{\nu_{3},+z} \\ + e^{i\frac{(\bar{\phi}_{1}+\bar{\phi}_{2})+(\omega_{1}+\omega_{2,3})t}{2}} U_{l 3}     \Big(  e^{-i\frac{\beta}{2}}\cos\frac{\alpha}{2}\sin\frac{\theta_{1}}{2}\sin\frac{\theta_{2}}{2} - e^{i\frac{\beta}{2}}\sin\frac{\alpha}{2}\cos\frac{\theta_{1}}{2}\cos\frac{\theta_{2}}{2}      \Big)\ket{-z}\ket{\nu_{3},-z}.
\end{gathered}
\end{equation}
During time evolution from $t=0$ to $t=\tau$, the total phase $\Phi_{T}$ and dynamical phase $\Phi_{D}$ are respectively 
\begin{equation}
\begin{gathered}
     \Phi_{T}={\rm arg}\bra{\Psi(0)}\ket{\Psi(\tau)}=   {\rm arg}\Big\{ \sum_{j=1}^{3}|U_{l j}|^{2}\Big[ \cos\frac{\omega_{1}\tau}{2}\cos\frac{\omega_{2,j}\tau}{2} -\sin\frac{\omega_{1}\tau}{2}\sin\frac{\omega_{2,j}\tau}{2}\cos\theta_{1}\cos\theta_{2} 
     \\ +8\sin\frac{\omega_{1}\tau}{2}\sin\frac{\omega_{2,j}\tau}{2} \cos\beta \cos\frac{\alpha}{2}\cos\frac{\theta_{1}}{2}\cos\frac{\theta_{2}}{2}     \sin\frac{\alpha}{2}\sin\frac{\theta_{1}}{2}\sin\frac{\theta_{2}}{2}\\
     -i \Big(\sin\frac{\omega_{1}\tau}{2}\cos\frac{\omega_{2,j}\tau}{2}\cos\alpha\cos\theta_{1}+\cos\frac{\omega_{1}\tau}{2}\sin\frac{\omega_{2,j}\tau}{2}\cos\alpha\cos\theta_{2}\Big)\Big]\Big\},
\end{gathered}
\end{equation}
\begin{equation}
\begin{gathered}
    \Phi_{D}=-i\int_{0}^{\tau}\bra{\Psi(t)}\ket{\dot{\Psi}(t)}dt=-\sum_{j=1}^{3} |U_{l j}|^{2}\Big[ \frac{\omega_{1}\tau}{2}\cos\alpha\cos\theta_{1}+\frac{\omega_{2,j}\tau}{2}\cos\alpha\cos\theta_{2}\Big].
\end{gathered}
\end{equation}
Hence the noncyclic geometric phase obtained by an entangled pair consisting of one charged lepton and one neutrino initially in its $l$-flavor eigenstate during the time evolution from $t=0$ to $t=\tau$ is 
\begin{equation}
\begin{gathered}
    \Phi_{G}=\Phi_{T}-\Phi_{D}\\
    ={\rm arg}\Big\{ \sum_{j=1}^{3}|U_{l j}|^{2}\Big[ \cos\frac{\omega_{1}\tau}{2}\cos\frac{\omega_{2,j}\tau}{2} -\sin\frac{\omega_{1}\tau}{2}\sin\frac{\omega_{2,j}\tau}{2}\cos\theta_{1}\cos\theta_{2} 
     \\ +8\sin\frac{\omega_{1}\tau}{2}\sin\frac{\omega_{2,j}\tau}{2} \cos\beta \cos\frac{\alpha}{2}\cos\frac{\theta_{1}}{2}\cos\frac{\theta_{2}}{2}     \sin\frac{\alpha}{2}\sin\frac{\theta_{1}}{2}\sin\frac{\theta_{2}}{2}\\
     -i \Big(\sin\frac{\omega_{1}\tau}{2}\cos\frac{\omega_{2,j}\tau}{2}\cos\alpha\cos\theta_{1}+\cos\frac{\omega_{1}\tau}{2}\sin\frac{\omega_{2,j}\tau}{2}\cos\alpha\cos\theta_{2}\Big)\Big]\Big\}\\
    + \sum_{j=1}^{3} |U_{l j}|^{2}\Big[ \frac{\omega_{1}\tau}{2}\cos\alpha\cos\theta_{1}+\frac{\omega_{2,j}\tau}{2}\cos\alpha\cos\theta_{2}\Big].
\end{gathered}
\end{equation}
It is easy to notice that only squares of magnitudes of entries in PMNS matrix appear in the above formula. If massive neutrinos are Majorana particles, we need simply substitute $\omega_{2,j}=0$ for all $j=1,2,3$, i.e.,
\begin{equation}
\begin{gathered}
   \Phi_{G}
    =-\arctan\Big( \tan\frac{\omega_{1}\tau}{2}\cos\alpha\cos\theta_{1}\Big)
    + \frac{\omega_{1}\tau}{2}\cos\alpha\cos\theta_{1}.
\end{gathered}
\end{equation}
It is different from the noncyclic geometric phase obtained by a single charged lepton in constant magnetic field by a factor $\cos\alpha$. Although massless neutrinos and massive Majorana neutrinos have vanishing Larmor frequency, their entanglement with the charged lepton is still reflected in the geometric phase of the whole system.\\
In the Standard Model, we know that charged pseudoscalar mesons can produce a charged lepton-neutrino pair via purely leptonic decay intermediated by a virtual $W$ boson in tree level. \cite{PDG} The ejected charged lepton and neutrino are maximally entangled. In the previous analysis, we have implicitly made an assumption that the interactions between two subsystems are all negligible. This assumption works well in the leptonic decay of charged pseudoscalar mesons. First, the neutrino is electrically neutral, thus the electric force between two subsystems can be safely ignored. Second, in the center-of-momentum frame, once the leptonic decay happens the charged lepton and the neutrino will propagate in opposite directions with the spatial separation increasing rapidly, which is accompanied by a rapidly decreasing strength coefficient in the spin-spin interaction Hamiltonian. Third, the gravitational interaction is obviously nonsignificant due to the smallness of masses. At the present stage, we are still expecting hints regarding the Dirac/Majorana nature of massive neutrinos from neutrinoless double beta decay ($0\nu\beta\beta$), which violates the conservation of lepton number and can happen only if massive neutrinos are Majorana particles. \cite{petcov}\cite{0nu} The charged pseudoscalar mesons can be candidates of sources in the future investigation of neutrino nature once the problem of detection of geometric phases related to neutrinos is solved.

\section{CONCLUSION}
We have calculated the noncyclic geometric phase for entangled charged lepton-neutrino pair in the constant magnetic field, neglecting the interactions between subsystems. This geometric phase has nontrivial dependence on neutrino masses and can serve as a tool to determine the Dirac/Majorana nature of massive neutrinos.

\section*{ACKNOWLEDGEMENTS}
This work is supported by National University of Singapore Research Scholarship.


\end{document}